\documentclass[twocolumn,showpacs,preprintnumbers,amsmath,amssymb]{revtex4}

\usepackage{graphicx}
\usepackage{dcolumn}
\usepackage{bm}

\begin{document}


\title{Mean-field behavior of the negative-weight 
percolation model on random regular graphs}
\author{Oliver Melchert}
\email{oliver.melchert@uni-oldenburg.de} 
\author{Alexander K. Hartmann}
\affiliation{Institute of Physics, University of Oldenburg, 26111 Oldenburg, 
Germany}
\author{Marc M\'ezard}
\affiliation{Laboratoire de Physique Th\'eorique et Modeles Statistiques,
Universit\'e de Paris Sud,
91405 Orsay, France}
\date{\today}


\begin{abstract}
We investigate both analytically and numerically 
the ensemble of minimum-weight loops and paths in the negative-weight
percolation model on random graphs with fixed connectivity
and bimodal weight distribution. 
This allows us to study the mean-field behavior of this model.
The analytical study is based on a conjectured equivalence
with the problem of self-avoiding walks in a random medium. 
The numerical study is based
on a mapping to a standard minimum-weight matching problem for which fast
algorithms exist. Both approaches yield results which are in
agreement, on the location of the phase transition, on the value of
critical exponents, and on the absence of any sizeable indications of
a glass phase. By these results, the previously conjectured
upper critical dimension of $d_u=6$ is confirmed.
\end{abstract} 

\pacs{}
\maketitle

\section{Introduction  \label{sect:introduction}}

The statistical properties of lattice-path models on graphs,
equipped with quenched disorder, have experienced much attention during the 
last decades.
They have proven to be useful in order to describe line-like quantities as, e.g.,
linear polymers in disordered media \cite{kardar1987,derrida1990,grassberger1993,buldyrev2006}, 
vortices in  high $T_c$ superconductivity \cite{pfeiffer2002,pfeiffer2003},
cosmic strings in the early universe \cite{vachaspati1984,scherrer1986,hindmarsch1995},
and domain-wall excitations in disordered systems such as $2d$ spin glasses \cite{cieplak1994,melchert2007} and 
the $2d$ solid-on-solid model \cite{schwarz2009}. 
The precise computation of these paths can often be formulated in terms
of a combinatorial optimization problem and hence might allow for the
application of exact optimization algorithms developed in computer science \cite{schwartz1998,rieger2003,SG2dReview2007}.
So as to analyze the statistical properties of these lattice path models, 
geometric observables and scaling concepts similar to those used in 
percolation theory \cite{stauffer1979,stauffer1994} or other ``string''-bearing
models \cite{allega1990,austin1994} are often applicable. 

This paper studies the 
\emph{negative-weight percolation} 
(NWP)  problem \cite{melchert2008,apolo2009,melchert2010a}, 
wherein one considers a graph where sites are joined by undirected edges. 
Weights are
assigned to the edges, representing quenched random variables drawn from 
a distribution that allows for edge weights of either sign. 
The details of the weight distribution are further controlled by a tunable 
disorder parameter.
We are interested in    
configurations consisting of one path and a set of loops, i.e.\ closed paths on the graph, 
such that the total sum of the weights assigned to the edges
that build up the path and the loops attains a 
 minimum. 
We will study here the mean-field behavior of this model
by means of numerical simulations 
and analytical studies on random regular
graphs. For the numerical studies,
one could use
standard sampling approaches like Monte Carlo simulations
but they  exhibit the usual equilibration problems. 
Fortunately, one can map NWP to the standard minimum-weight
perfect matching optimization
problem, as  outlined below in sect.\ \ref{sect:model} in 
more detail. This mapping
allows to apply exact polynomial-time-running algorithms.
Thus,
 large instances can be solved.
As an additional optimization constraint we impose the condition that 
the string-like observables (referring to both, the loops and the path) are not allowed to intersect. 
Consequently, since a string-like observable does neither intersect with itself nor 
with other strings in its neighborhood, it exhibits an ``excluded volume''
quite similar to usual self avoiding walks (SAWs) \cite{stauffer1994}.
 We shall develop in this paper the analogy
  between SAWs and NWP, and use some of the recently developed tools in
  the studies of polymers on random graphs in order to derive an
  analytical study of the NWP problem.

As presented here, the NWP problem is a theoretical model of intrinsic
interest.
As an example that makes use of the NWP problem statement one can imagine an 
agent that travels on a graph. While traversing
an edge, the agent either needs to pay some resource (signified by a 
positive edge weight) or he is able, once per edge,
 to harvest some resource (signified by a negative
edge weight). Now, if the intention of the agent is to gain as much resources as possible, 
the paths/loops obtained in the context of the NWP problem can serve as a guide to find
routes along which the agent might move so as to optimize his yield.
Further, the 2d variant of the NWP problem is interesting from a technical
and algorithmic point of view.
As regards this, the problem of finding ground-state spin 
configurations for
the $2d$ random-bond Ising model, including the canonical Ising spin glass, on
a planar triangular lattice can be mapped to the 2d NWP problem on a
honeycomb lattice \cite{melchertThesis2009,melchert2011} (Analogous to the approach presented in
\cite{thomas2007,pardella2008}, the basic idea of this mapping is to compute
a transition graph that mediates the transformation from an initially chosen
reference spin configuration to a proper GS).
Moreover, paths that include negative edge-weights also appear in the context of 
domain wall excitations in $2d$ random bond Ising systems \cite{melchert2007,melchert2009}.
Thus, the NWP problem might serve to gain insight concerning the behavior
of more realistic disordered systems.

Previous studies \cite{melchert2008,apolo2009,melchert2010a} of the NWP model considered
regular lattice graphs in dimension $d$ with periodic boundary conditions. 
As a pivotal observation it was found that, as a function of the disorder parameter,  
the NWP model features a disorder-driven, geometric phase transition.
This transition leads from a phase characterized by only ``small'' loops to a phase that also
features ``large'' loops that span the entire lattice along at least one direction.
Regarding these two phases and in the limit of large system sizes, there is a particular 
value of the disorder parameter at which percolating (i.e.\ system spanning) loops appear 
for the first time \cite{melchert2008}.     
Previously, we have investigated the NWP phenomenon for $2d$ lattice graphs \cite{melchert2008}
using finite-size scaling (FSS) analyses, where we characterized the underlying transition by 
means of a whole set of critical exponents. 
Considering different disorder distributions and lattice geometries, the exponents were found 
to be universal in $2d$ and clearly distinct from those describing other percolation phenomena.
In a subsequent study, we performed further simulations for the NWP model on hypercubic lattice 
graphs in dimensions $d\!=\!2$ through $7$ \cite{melchert2010a}, where we aimed to assess the
upper critical dimeison $d_u$ of the NWP model.
As a fundamental observable that provides information on whether the upper critical 
dimension $d_u$ is reached, we monitored the fractal dimension $d_f$ of the loops. 
The fractal dimension can be defined from the scaling of the average length $\langle \ell \rangle$
of the percolating loops as a function of the linear extension $L$ of the lattice graphs, 
according to $\langle \ell \rangle\!\sim\!L^{d_f}$.
For $d\!\geq\!d_u$ one might expect to observe $d_f\!=\!2$, which signifies the mean-field limit for 
self-avoiding lattice curves. This means, the ``excluded volume'' effect mentioned earlier 
becomes irrelevant and the loops exhibit the same scaling as ordinary random walks. 
From that study, considering regular $d$-dimensional lattice graphs, we found evidence for an 
upper critical dimension $d_u\!=\!6$ for the NWP phenomenon.

%
\begin{figure}[t!]
\centerline{
\includegraphics[width=0.8\linewidth]{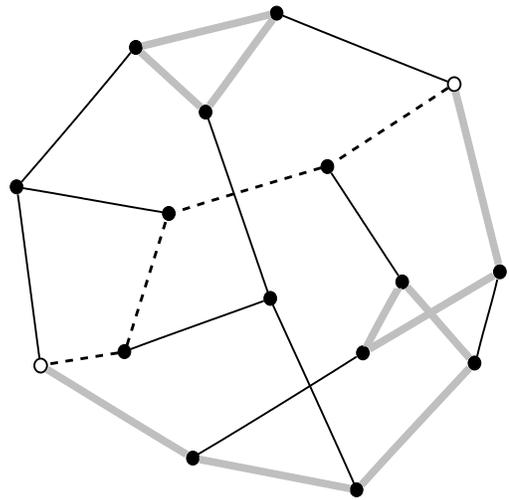}}
\caption{Example of a $r$-regular random graph with $N\!=\!16$ nodes and 
fixed degree $r\!=\!3$.
The depicted graph has a diameter $R_{N,r}\!=\!4$, where the diameter is 
the longest among all shortest paths (measured in node-to-node hops).
For a given realization of the quenched disorder, the negative-weight percolation
algorithm might yield the 
path+loop configuration illustrated by 
means of the bold gray edges. Therein, a path is forced to join 
two distinguished nodes (open circles) that are maximally seperated, i.e.\
have distance $R_{N,r}$. The dashed path corresponds to a shortest path joining the
two distinguished nodes. 
\label{fig1ab}}
\end{figure}  

To resume, 
previous works have focused on the critical properties of the NWP model 
on regular lattice graphs, where the upper critical dimension can be 
defined as the smallest dimension for which the critical exponents 
take their mean field values.
Here, we perform simulations and analytical
calculations for the NWP problem on random graphs with fixed connectivity 
(see discussion below), where one has direct access to the 
mean-field exponents that 
describe the transition. These 
studies provide further support 
for the result $d_u\!=\!6$ obtained previously \cite{melchert2010a}.
Furthermore, we examine the structure of the energy
  landscape of this problem by studying the existence of
low-energy excitations involving a finite fraction of the system. 
The numerical simulations presented here are carried out on the ensemble of 
random regular graphs with a fixed node degree. 
Due to the absence of, say, a regular lattice geometry that allows for 
a clear cut definition of spanning lengths of loops (as it was possible
in previous studies),
we found it more beneficial to focus on the scaling properties of minimum 
weight paths instead. 
In this regard, if we select two nodes on a random graph we can compare
the properties of the minimum weight path to the shortest path that connect
the two distinguished nodes. In the context of the presented study, this 
will provide means to assess the scaling properties of the paths.
Further, note that in a previous study \cite{melchert2008} we verified that loops and paths
in the NWP problem give rise to identical scaling
properties.

The remainder of the presented article is organized as follows.
In section \ref{sect:model}, we introduce the model in more detail and 
we outline the algorithm used to compute the minimum weight configurations 
of loops. In section \ref{sect:results}, we present the results of 
our numerical simulations. In section \ref{sect:cavity},
we explain the analytical cavity approach through a mapping to
  SAWs, and we  show the corresponding
results. Finally,
in section \ref{sect:conclusions} we 
conclude with a summary. 
Note that an extensive summary of this paper
is available at the \emph{papercore database} \cite{papercore}.


\section{Model and algorithm  \label{sect:model}}

In the remainder of this article we consider $r$-regular random graphs ($r$-RRGs) \mbox{$G_{N,r}\!=\!(V,E)_{N,r}$}, wherein
$V$ is a set of $N$ nodes $i\!\in\!V$, having fixed degree ${\rm deg}(i)\!=\!r$.
Accordingly, $E$ is a set of $rN/2$ edges $e_{ij}\!=\!\{i,j\}$, $i,j\!\in\!V$, randomly drawn 
from $V^{(2)}$ with the restriction that each node has exactly $r$ neighbors.
Further, the distance $d_{ij}$ between two nodes $i,j\!\in\!V$ is the number of steps in the 
shortest (i.e.\ minimum-length) path joining both nodes.
If the respective nodes cannot be joined by a path, the distance is taken as infinite.
Then, the diameter $R_{N,r}$ of an instance of $G_{N,r}$ is the largest finite distance 
within the given graph.
Further, two nodes $i,j\!\in\!V$ satisfying $d_{ij}\!=\!R_{N,r}$ are said to be maximally separated.
Fig.\ \ref{fig1ab} illustrates an istance of $G_{N,r}$ with $N\!=\!16$ and $r\!=\!3$. 
At this point we would like to point out that there exists an intricate relation between 
the number of nodes and the diameter of $r$-RRGs. 
In this regard, early estimates for fixed $r\!\geq\!3$ yield the result 
that there is a \mbox{$r$-RRG}
of diameter at most 
$R_{N,r}\!=\!\lceil \log_{r-1}( (2+\epsilon) N \log(N)/r) \rceil \!+\! 1$ 
(where $\epsilon\!>\!0$) 
\cite{bollobas1982,bollobas2001}.
Later, it was indicated that the diameter of $3$-RRGs scales as 
$R_{N,3}\!=\!1.4722 \log(N)\!+\!O(1)$ \cite{jerrum1984}. Further, for $r\!=\!4$ the relation 
$R_{N,4}\!=\!0.9083 \log(N)\!+\!O(1)$ was found \cite{jerrum1984}. These results also agree with the expression
$R_N\!=\!c \log(N)\!+\!O(\log(N))$ for the diameter of sparse random graphs with specified degree sequence 
\cite{fernholz2007}.  
Subsequently we fix $r\!=\!3$ and substitute $R_{N,r}\equiv R_{N}$ 
and $G_{N,r}\equiv G$.

We further assign a weight $\omega_{ij}$ to each edge contained in $E$, 
representing quenched random variables that introduce 
additional disorder to the random graph ensemble.
In the presented article we consider weights
that are drawn from the bimodal disorder distribution
\begin{equation}
  \label{eq:def:weights}
  P(\omega)\!=\! \rho \delta(\omega+1) + (1-\rho)\delta(\omega-1)\,. 
\end{equation}
The disorder parameter $\rho$ therein allows to 
adjust the fraction of negative edge weights on the graph.
Note that this disorder distribution 
explicitly allows for loops $\mathcal{L}$ with a negative total weight
$\omega_{\mathcal{L}}\!=\!\sum_{\{i,j\}\in\mathcal{L}}\omega_{ij}$.
To support intuition: For any nonzero value of the disorder parameter $\rho$, a sufficiently
large graph will exhibit at least ``small'' loops that have negative weight.

The NWP problem statement then reads as follows:
Given $G$ together with a realization of the disorder,
determine a set $\mathcal{C}$ of loops such that the
configurational energy, defined as the sum of all the loop-weights
$\mathcal{E}\!=\!\sum_{\mathcal{L} \in \mathcal{C}} \omega_{\mathcal{L}}$, is
minimized.  As further optimization constraint, the loops are not
allowed to intersect and generally, the  weight of an individual loop
is smaller than zero. Note that $\mathcal{C}$ may also be empty 
(clearly this is the case for $\rho\!=\!0$).
Clearly, the configurational energy $\mathcal{E}$ is the quantity subject to
optimization and the result of the optimization procedure is a set of
loops $\mathcal{C}$, obtained using an appropriate transformation of
the original graph as detailed in \cite{ahuja1993}.  So as to identify
the edges that constitute the loops for a particular instance  of the
disorder, we can benefit from a relation between minimum-weight paths 
(and loops) on $G$ and minimum-weight perfect matchings (MWPM) 
\cite{cook1999,opt-phys2001,melchertThesis2009} on the transformed graph.
Here, we give a brief description of the algorithmic procedure that yields a 
minimum-weight set of loops for a given realization of the disorder. 
Fig.\ \ref{fig2abcd} illustrates the 3 basic steps, detailed below:

\begin{figure}[t!]
\centerline{\includegraphics[width=1.0\linewidth]{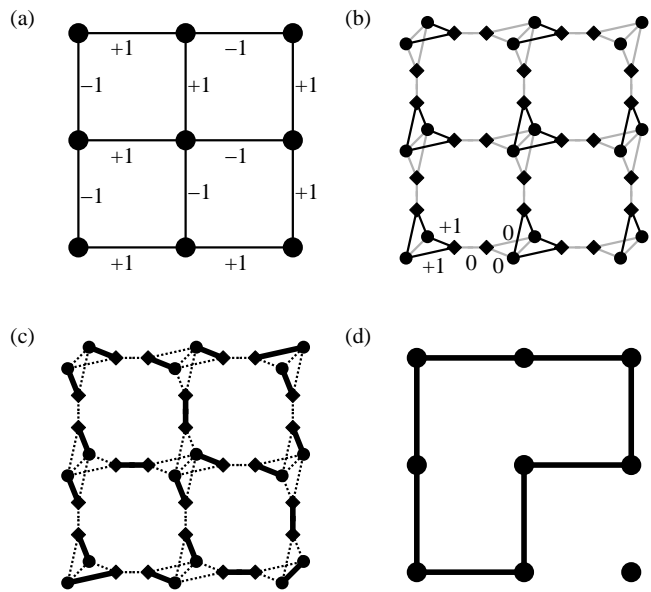}}
\caption{Illustration of the algorithmic procedure:
(a) original lattice $G$ with edge weights, 
(b) auxiliary graph $G_{\rm A}$ with proper weight assignment. Black 
edges carry the same weight as the respective edge in the original
graph and grey edges carry zero weight,
(c) minimum-weight perfect matching (MWPM) $M$: bold edges are matched 
and dashed edges are unmatched, and
(d) loop configuration (bold edges) that corresponds to the MWPM 
depicted in (c).
\label{fig2abcd}}
\end{figure}  

(1) each edge, joining adjacent sites on the original graph $G$,  is
replaced by a path of 3 edges.  Therefore, 2  ``additional'' sites
have to be introduced for each edge in $E$.  Therein, one of
the two edges connecting an additional site to an original site gets
the same weight as the corresponding edge in $G$. The remaining  two
edges get zero weight.  The original sites $i\in V$ are then
``duplicated'',  i.e. $i \rightarrow i_{1}, i_{2}$, along with all
their incident edges and the corresponding weights. 
 For each of these pairs of duplicated sites,
one additional  edge $\{i_1,i_2\}$ with zero weight is added that
connects the two sites $i_1$ and $i_2$.  The resulting auxiliary graph
$G_{{\rm A}}=(V_{{\rm A}},E_{{\rm A}})$  is shown
in Fig.\ \ref{fig2abcd}(b), where additional sites appear as squares and
duplicated  sites as circles. Fig.\ \ref{fig2abcd}(b) also illustrates
the weight  assignment on the transformed graph $G_{{\rm A}}$.  Note
that while the original graph (Fig.\ \ref{fig2abcd}(a)) is symmetric, the
transformed graph (Fig.\ \ref{fig2abcd}(b)) is not. This is due to the
details of the mapping procedure and the particular weight assignment
we have chosen.  A more extensive description of the mapping can be
found in \cite{melchert2007}.

(2) a MWPM on the auxiliary graph is determined via exact
combinatorial-optimization algorithms \cite{comment_cookrohe}.  A MWPM
is a minimum-weighted subset $M$ of $E_{\rm A}$, such that
each site  contained in $V_{\rm A}$ is met by precisely one
edge in $M$.  This is illustrated in Fig.\ \ref{fig2abcd}(c), where the
solid edges  represent $M$ for the given weight assignment. The dashed
edges are  not matched.  Due to construction of the auxiliary graph,

(3) finally it is possible to find a relation between the matched
edges $M$  on $G_{\rm A}$ and a configuration of negative-weighted
loops  $\mathcal{C}$ on $G$ by  tracing back the steps of the
transformation (1). As regards this, note that each edge  contained
in $M$ that connects an additional site (square) to a duplicated  site
(circle) corresponds to an edge on $G$ that is part of a loop, see
Fig.\ \ref{fig2abcd}(d). More precisely, there are always two such edges
in $M$  that correspond to one loop segment on $G$. All the edges in
$M$ that connect like  sites (i.e.\ duplicated-duplicated, or
additional-additional)  carry zero weight and do not contribute to a
loop on $G$.  Once the set $\mathcal{C}$ of loops is found, a
depth-first search \cite{ahuja1993,opt-phys2001} can  be used to
identify the loop set $\mathcal{C}$ and to determine the geometric 
properties of the individual loops. For the weight assignment illustrated 
in Fig.\ \ref{fig2abcd}(a), there is only one negative weighted loop with
$\omega_{\mathcal L}\!=\!-2$ and length $\ell=8$.

Note that the result of the calculation is a collection $\mathcal{C}$
of loops such that the total loop weight, and consequently the
configurational energy $\mathcal{E}$, is minimized. 
Hence, one obtains a global collective optimum of the system.  
Obviously, all loops that contribute to $\mathcal{C}$ possess a negative weight.  
Also note that the choice of the weight assignment in step (1) is not unique, 
i.e.\ there are different possibilities to choose a weight assignment
that all result in equivalent sets of matched edges on the transformed
lattice, corresponding to the minimum-weight collection of loops on
the original lattice. Some of these weight assignments result in a more
symmetric  transformed graph, see e.g. \cite{ahuja1993}. However, this
is only a technical issue that does not affect the resulting loop
configuration. Albeit the  transformed graph is not symmetric, the
resulting graph (Fig.\ \ref{fig2abcd}(d)) is again symmetric.

\begin{figure}[t!]
\centerline{\includegraphics[width=1.0\linewidth]{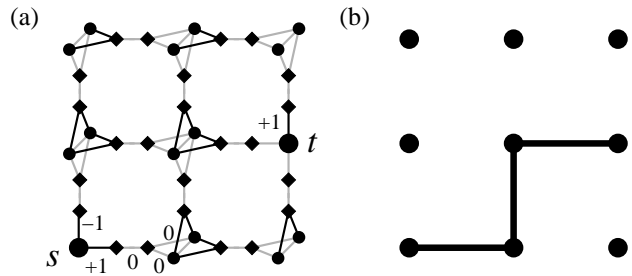}}
\caption{Illustration of the algorithmic procedure to obtain a 
minimum weight $s$-$t$-path:
(a) auxiliary graph $G_{\rm A}$, where the mapping is modified to 
induce a minimum weight path that connects nodes $s$ and $t$.
Black edges carry the same weight as the respective edge in the original
graph and grey edges carry zero weight,
(b) minimum weight path (bold edges) that corresponds to a MWPM 
on $G_{\rm A}$. For the particular example illustrated here, the
path has weight $\omega_{p}=-1$ and there
are no loops in addition to the path.
\label{fig:path_mapping}}
\end{figure}  
Note that the above description explains how to obtain a set of loops only.
If one aims to compute an additional minimum weight path that connects two
nodes, say $s$ and $t$, on the graph, the transformation procedure for these
two particular nodes will look slightly different: the duplication 
step introduced in step (1) will be skipped for nodes $s$ and $t$, see 
Fig.\ \ref{fig:path_mapping}(a).
Computing a MWPM for the resulting graph will then yield a minimum weight 
path that connects nodes $s$ and $t$ together with a set of loops (the set
might be empty) as explained in steps (2) and (3) above. 
This is illustrated in Fig.\ \ref{fig:path_mapping}(b), where for the same
weight assignment as in Fig.\ \ref{fig2abcd}(a), a minimum weight path 
is computed. 
For this illustrational purpose, a small $2d$ lattice graph with
free BCs was chosen. The algorithmic procedure extends 
to $r$-regular random graphs in a straightforward manner.

In the following we will use the procedure outlined above so as to 
investigate the NWP phenomenon on $3$-regular random graphs.


\section{Results \label{sect:results}}
In the following we will present the results obtained from 
numerical simulations of negative weighted loops and paths 
on $3$-regular random graphs.

\subsection{Minimum-weight paths (MWPs)} 
\begin{figure}[t!]
\centerline{\includegraphics[width=1.\linewidth]{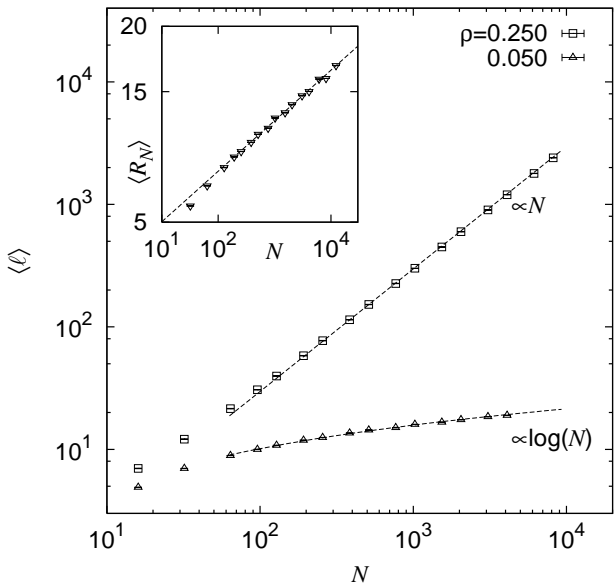}}
\caption{Scaling behavior of the minimum-weight path length
on RRGs respecting a bimodal distribution of the edge-weights.
The main plot shows the scaling of the average path length $\langle\ell\rangle$ 
at the values $\rho\!=\!0.05$ and $0.25$ of the disorder parameter,
yielding the scaling behavior $\sim\!\log(N)$ and $\sim\!N$, respectively.
The inset illustrates the scaling of the average RRG diameter as
$\langle R_N\rangle\!\sim\!1.68(5)\log(N)+1.2(5)$ (note that only the $x$-axis is scaled
logarithmically).
\label{fig3}}
\end{figure}

Once a particular instance of $G$ is constructed and so as to get a grip on the node-to-node 
distances $d_{ij}$ for $i,j\!\in\!V$, we traverse the graph using a depth-first search (DFS) \cite{clrs2001}.
Invoked to compute the distances from a particular source node to all other nodes in the graph, 
the DFS terminates in $O(N)$ time.
As soon as the DFS is completed for each node in the graph, the diameter of the particular random 
graph is easily obtained as the largest finite distance in $O(N^2)$ time.
Then we compute a minimum weight path on the graph that joins two distinguished nodes 
$s,t\!\in\!V$, satisfying $d_{st}\!=\!R_N$. 
For the range of system sizes considered here, and in qualitative agreement with the $R_N-N$-relation
given above, we found the approximate scaling $R_N\!=\!1.68(5) \log(N) + 1.2(5)$, see inset of Fig.\ \ref{fig3}. 
As regards this, note that for large graph sizes the diameter does not change much. 
An increase of $N\!=\!10^3 \to 10^4$ only leads to an increase of $R_N\!=\!13 \to 16.5$. 
For the scaling behavior of the average length $\langle \ell \rangle$ of the minimum-weight paths
we can expect that 
\begin{eqnarray}
\langle \ell \rangle \sim \left\{\begin{array}{ll}
         R_N\!\sim\!\log(N) & \mbox{for $\rho\!\to\!0$};\\
         N   & \mbox{for $\rho\!\to\!1$}. 
\end{array} \right.\label{eq:scalingLimits_l}
\end{eqnarray}
This is based on the intuition that as $\rho\to 0$, an increasing path length will lead to an increasing
path weight. As a result, a minimum weight path that connects two nodes will most likely coincide 
with the shortest path, i.e.\ a path using the minimum number of edges. In contrast, as $\rho\to1$
it becomes feasible to take large detours (relative to the shortest path) so as to traverse many edges
with negative weight. Consequently, an increasing path length will result in a decreasing path weight. 

Within our 
simulations we find $\langle \ell \rangle\!\sim\!R_N^{1.05(1)}$ at $\rho\!=\!0.04$
and $\langle \ell \rangle \sim N^{0.998(2)}$ at $\rho\!=\!0.25$, in agreement with Eq.\ \ref{eq:scalingLimits_l}, 
see Fig.\ \ref{fig3}.
Here, investigating MWPs on RRGs allows for a direct study of paths in the mean-field setup of
the negative-weight percolation model.
Now, since $R_N$ is the distance spanned by the MWPs upon construction, we can expect to find 
the asymptocic scaling behavior $\langle \ell \rangle \sim R_N^2$ at the critical point $\rho_c$ of the setup.
This would indicate the same statistical properties as for usual random walks, wherein the directions
of consecutive steps along the walk are completely uncorrelated.

\subsection{Bimodal disorder -- Path weight}

\begin{figure}[t!]
\centerline{\includegraphics[width=1.\linewidth]{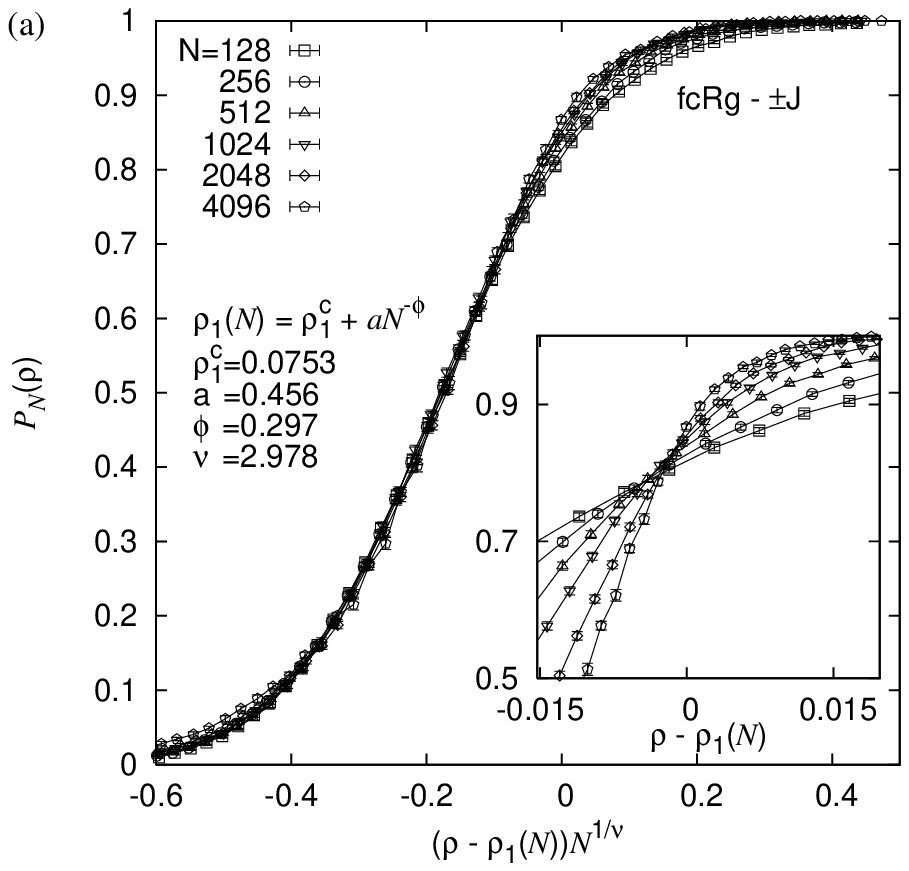}}
\centerline{\includegraphics[width=1.\linewidth]{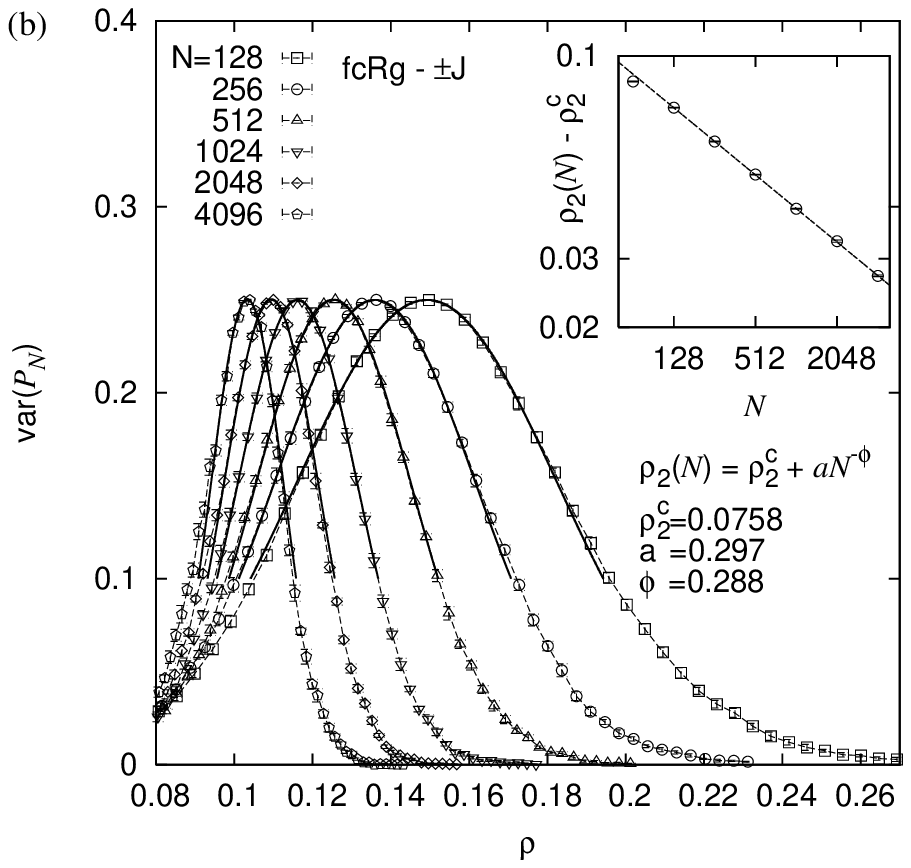}}
\caption{Results for minimum weight paths on RRGs respecting
a bimodal distribution of the edge-weights.
(a) Probability $P_N^\omega\!\equiv\!P_N(p_\omega\!\leq\!0)$ 
that a path has negative/zero weight. The main plot shows 
the rescaled data that also accounts for a drift in the
effective critical point $\rho_1(N)$.
The inset illustrates the crossing point of the data 
sets after correcting for the drift.
(b) Variance of $P_N^\omega$, related to the finite-size 
suszeptibility $C_N^\omega$ via $C_N^\omega=N {\rm var}(P_N^\omega)$.
The main plot shows the approximation of the 
${\rm var}(P_N^\omega)$-peaks by means of fits to a gaussian function. 
The inset illustrates the the scaling of the peak locations as 
$\rho_c(N) = \rho_c^\infty + a N^{-\phi}$.
\label{fig2ab}}
\end{figure}

First, we consider the probability $P^{\omega}_N\!\equiv\!P_N(p_{\omega}\!\leq\!0)$ that
the path weight $p_{\omega}$ is smaller/equal to zero. For each realization 
of the bond disorder, $P^{\omega}_N$ is either $0/1$. Hence, for the average value
$\langle P^{\omega}_N \rangle$ we expect a scaling behavior similar to the spanning
probability in percolation theory, i.e.\
$\langle P^{\omega}_N \rangle \sim f_0[(\rho-\rho_c^\infty) N^{1/\nu^*}]$,
wherein $f_0[\cdot]$ is a size-independent scaling function.
The value of $\rho_c^\infty$ signifies the location of the critical point, above which,
in the limit $N\!\rightarrow\!\infty$, a path with $p_\omega\!\leq\!0$ appears for the 
first time. 
From our simulations we find that with increasing $N$ this crossing point shifts towards 
smaller values of $\rho$. This might be attributed to the finite-size of the studied
systems and signals corrections to the scaling behavior. 
So as to account for these corrections to scaling, one may
consider an effective scaling expression of the form
\begin{eqnarray}
\langle P^{\omega}_N \rangle \sim f_1[(\rho-\rho_{1}(N)) N^{1/\nu^*_{1}}], \label{eq:pmProbApprox2}
\end{eqnarray} 
wherein $\rho_{1}(N)\!=\! \rho_{1}^c + a N ^{-\phi_{1}}$. 
The latter effective scaling form implies $4$ adjustable parameters and accounts for 
a shift in the effective critical point $\rho_1(N)$.
Considering Eq.\ \ref{eq:pmProbApprox2}, a best data collapse of the 
curves for $N\!\geq\!1024$ then yields the parameters
\begin{align*}
\rho^c_{1}      &= 0.075(1)\\
\phi_{1}        &= 0.297(7) \\
\nu^*_1         &= 3.0(1) 
\end{align*}
with $a\!=\!O(1)$, see Fig.\ \ref{fig2ab}(a).

As an alternative, we can estimate the critical point from the location of 
the maxima of the related finite-size suszeptibilities 
\begin{eqnarray}
C_N^\omega = N \langle (P_N^\omega-\langle P_N^\omega \rangle)^2\rangle= N {\rm var}(P_N^\omega). \label{eq:pmSusz}
\end{eqnarray}
In this regard, if, for a given system size $N$, the maximum of ${\rm var}(P_N^\omega)$ (or similary $C_N^\omega$) 
is located at the effective critical point $\rho_{2}(N)$, we expect the 
sequence of $\rho_2(N)$ for increasing $N$ to approach an asymptotic value of
$\rho_{2}^c$ according to 
\begin{eqnarray}
\rho_{2}(N) = \rho^c_{2} + a N^{-\phi}. \label{eq:pmpEff}
\end{eqnarray}
Here, $\phi$ signifies an effective exponent that accounts for the corrections to scaling in 
a very basic manner. Note that if there are no corrections to scaling, it holds that $\phi\!=\!1/\nu^*$.
The locations $\rho_{2}(N)$ were obtained by a fit 
\cite{practicalGuide2009} of a gaussian function to the peak of 
${\rm var}(P_N^\omega)$, as illustrated in Fig.\ \ref{fig2ab}(b). The analyisis of $\rho_{2}(N)$
by means of Eq.\ \ref{eq:pmpEff} for $N\!\geq\!128$, shown in the inset of Fig.\ \ref{fig2ab}(b), then yields:
\begin{align*}
\rho^c_{2}      &= 0.0758(9)    & \chi^2/{\rm dof}   &= 0.54 \\
\phi            &= 0.288(5)     & {\rm dof}          &= 3 
\end{align*}
Finally, note that the effective scaling form considered here does not properly account 
for the corrections to the scaling behavior. Actually, a more sophisticated 
analysis should account for corrections to scaling via 
$\rho_c(N) = \rho_c^\infty + a (1+cN^{-\omega})N^{-1/\nu^*}$.
Unfortunately our data did not allow for a more elaborate analysis involving 5 free
parameters. However, if we apply the latter scaling form and fix the value 
of $\nu^*$ to the presumably correct value $\nu^*\!=\!3$ we find, again for $N\!\geq\!128$,
\begin{align*}
\rho_c^\infty   &= 0.077(6)     & \chi^2/{\rm dof}   &= 0.86 \\
\omega          &= 0.06(50)     & {\rm dof}          &= 2          
\end{align*}
wherein $a$ and $c$ are $O(1)$, and where the values of $\rho_c^\infty$ and $\omega$ are
in agreement with the values found from the analysis of Eq.\ \ref{eq:pmProbApprox2}. 
Note that the error associated with $\omega$ is rather large.

\begin{figure}[t!]
\centerline{
\includegraphics[width=1.\linewidth]{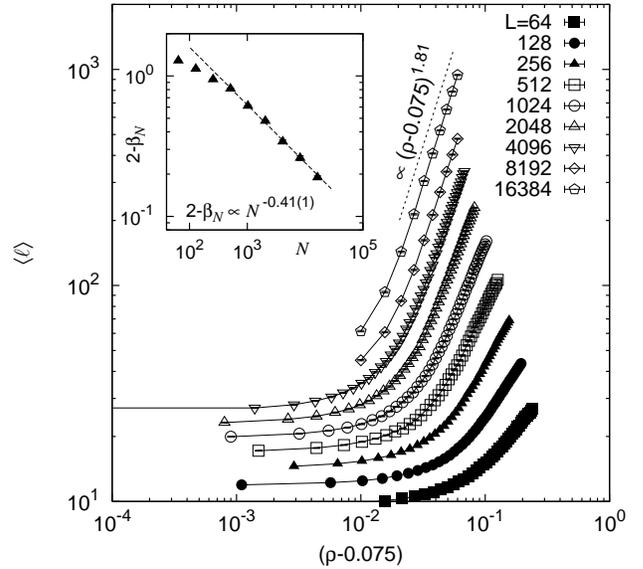}}
\caption{
Scaling of the average path length as a function of the distance to
the critical point $\rho_c=0.075(2)$ for different system sizes $N$.
Not too close to the critical point a power-law behavior is visible.
The dashed line shows an exemple of a  fit to a power law scaling
form $\sim (\rho -\rho_c)^{\beta_N}$, where at $N=16384$ we find $\beta_N=1.81(1)$.
The inset shows the obtained value $2-\beta_N$ as a function
of system size $N$. The line shows a power law
$\sim N^{-0.41}$. Please note the double logarithmic scales.
\label{fig:density:sim}
}
\end{figure}

\begin{figure}[t!]
\centerline{\includegraphics[width=1.\linewidth]{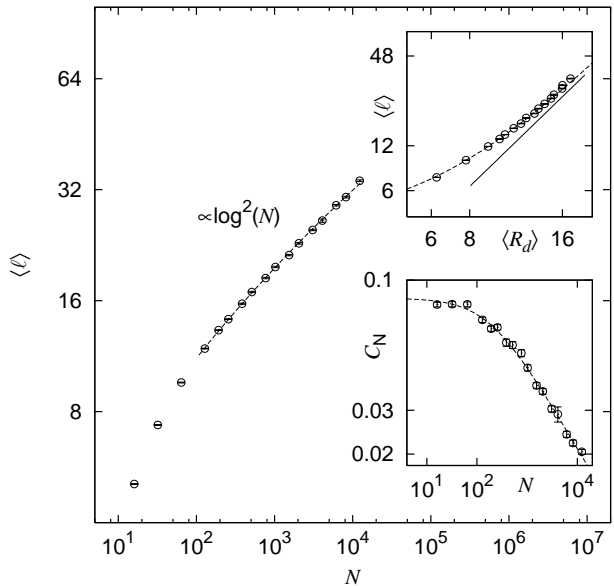}}
\caption{Scaling behavior of the minimum-weight path length
on RRGs respecting a bimodal distribution of the edge-weights at 
the critcal point $\rho_c\!=\!0.075(2)$.
The main plot shows the scaling of the average path length $\langle\ell\rangle$,
where the dashed line is a fit to the expression $\langle\ell\rangle\!\sim\!\log^2(N)(1+c/N^d)$.
The upper inset illustrates the scaling of $\langle\ell\rangle\!\sim\!R_N^2$ and the 
lower inset shows the scaling of the related finite-size susceptibilty at $\rho_c$.
\label{fig4}}
\end{figure}  


From the agreement among the extrapolated results we conclude with a critical point 
$\rho_c^\infty\!=\!0.075(2)$ and $\nu^*\!=\!3.0(1)$ in agreement with the value 
$\nu^*\!=\!d_c\nu\!=\!3$, that one could expect from the results for the 
finite-dimensional regular graphs.

Next, we determine the critical exponent $\beta$, describing the behavior
of the relative path length $\psi=\langle \ell \rangle/N$ 
as a function on the distance from the critical
point according to

\begin{equation}
  \label{eq:def:beta}
   \psi \sim (\rho -\rho_c)^{\beta}\,.
\end{equation}

In Fig.\ \ref{fig:density:sim}, the scaling of $N\psi$ according to
Eq.\ \ref{eq:def:beta} is shown for different values of $N$. The value
of the exponent obtained in the scaling region  depends on $N$,
i.e., $\beta=\beta_N$ (for the largest graph size considered, i.e.\ $N=16384$, we
find $\beta_N=1.81(1)$). The dependence is well compatible
with a finite-size scaling according to $\beta_N=\beta+aN^{-b}$
with $\beta=2$, $a=10(1)$, $b=-0.41(1)$, see inset of 
Fig.\ \ref{fig:density:sim}. Hence, this value of $\beta$ agrees
well with the numerical value $\beta=1.92(6)$ found \cite{melchert2010a}
for hypercubic lattices at the
presumed upper critical dimension $d=6$. It also agrees with the
analytical computation developed in Sect.\ref{sect:cavity}.

\subsection{Scaling at the critical point}

We further performed simulations to characterize the scaling behavior of the 
minimum-weight path length on RRGs respecting a bimodal distribution of the 
edge-weights at the critcal point $\rho_c\!=\!0.075(2)$, 
see Fig.\ \ref{fig4}.

For the average path length $\langle\ell\rangle$, we found a good agreement 
with the scaling expression 
\mbox{$\langle\ell\rangle\!\sim\!\log^{d_f}(N)+c_1$}. A fit to the data belonging
to $N\!>\!100$ yields $d_f\!=\!2.1(1)$ and $c_1\!=\!5.2(5)$
Considering the scaling expression
\mbox{$\langle\ell\rangle\!\sim\!\log^{d_f}(N)(1+c_1/N^{c_2})$}
(main plot of Fig.\ \ref{fig4}), and taking into account the data belonging to 
$N\!>\!100$ yields $d_f\!=\!2.3(3)$, $c_1\!=\!6(3)$ and $c_2\!=\!0.45(4)$.
Fixing $c_2\!=\!0.5$ improves the result to $d_f\!=\!2.1(1)$ and $c_1\!=\!6(1)$.
Hence, the data for the average path length at $\rho_c$ is in agreement with 
a scaling exponent $d_f\!=\!2$.

For the NWP problem in finite dimensions we introduced the order parameter $P_\infty = \ell/N$.
The respective finite-size suszeptibilities
$C_N = N {\rm var}(P_\infty) = N^{-1} {\rm var}(\ell)$
exhibit the scaling  $C_N\!\sim\!N^{\gamma/\nu}$  at $\rho_c$. Here, for the mean-field 
setup, the latter scaling is modified to $C_N\!\sim\!N^{\gamma/\nu^*}$. 
If we consider corrections to scaling according to $C_N\!\sim\!(N + c_1)^{\gamma/\nu^*}$ we 
yield the estimates $c_1\!\approx\!180$ and $\gamma/\nu^* \!=\!-0.34(2)$ (see lower inset of 
Fig. \ref{fig4}).
The more complicated scaling expression $C_N\!\sim\!N^{\gamma/\nu^*}(1+c_1/N^{c_2})$ yields qualitative similar results, i.e.\ 
$\gamma/\nu^*\!=\!-0.33(3)$, $c_1\!\approx\!35$ and $c_2\!=\!1.0(7)$.

\subsection{Excitations }
\label{sect:excitation}

Next, we want to examine, wether the model exhibits
low-energy excitations, which are of order of system size. This
would be a numerical evidence that NWP exhibits a complex energy
landscape, thus shows ``glassy'' behavior.

Here, the ``size'' of an excitation is defined relative to two 
ground state configurations of loops. I.e., for a given instance
of a $r$-regular random graph $G$ we compute two ground states (GSs) as 
follows:
\begin{itemize}
\item[(i)] Draw a realization of the (bimodal) bond disorder $\omega^{(1)}_{ij}$, $\{i,j\}\in E$, where $|\omega^{(1)}_{ij}|=1$. 
Then compute a minimum energy configuration of loops, i.e.\ GS 1, as explained earlier. The number of loop segments that 
build up the respective loops is refered to as $C_1$.
\item[(ii)] ``Perturb'' the realization of disorder according to 
$\omega^{(2)}_{ij} = \omega^{(1)}_{ij} + \epsilon n_{ij}$, where  $n_{ij}$ are independent and identically
distributed Gaussian random variables of zero mean and unit variance, and $\epsilon=1/100$. 
Compute the corresponding minimum energy configuration, i.e.\ GS 2, of loops and let $C_2$ denote the respective number of loop segments. 
\end{itemize}

\begin{figure}
\centerline{\includegraphics[width=1.\linewidth]{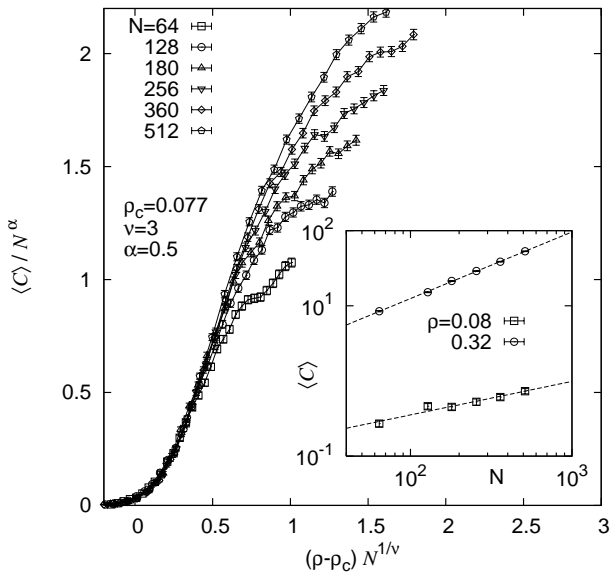}}
\caption{Scaling behavior of the average number $\langle C \rangle$ of 
bonds that change upon perturbing the ground state configuration 
of loops, as explained in the text. The main plot shows the 
scaled data and the inset shows the scaling of $\langle C \rangle$ 
as a function of the system size $N$. 
\label{fig5}}
\end{figure}

Now, consider the number $C=|C_2-C_1|$ of bonds that change by going from GS 1 to GS 2. 
The average value of $C$ exhibits a scaling of the form $\langle C \rangle = N^{\alpha} f[(\rho-\rho_c) N^{1/\nu^*} ]$,
where the choice $\rho_c\approx 0.077$, $\nu^*=3$ and $\alpha=0.5$ yields the data collapse illustrated in Fig.\ \ref{fig5}.
Further, the scaling behavior of $\langle C \rangle$ at
$\rho_c=0.077$  is not that clear. However, 
measurements at $\rho=0.08\approx\rho_c$ yield the scaling $\langle C \rangle \sim N^{0.46(4)}$ and 
at $\rho=0.32$ they yield $\langle C \rangle \sim N^{0.89(1)}$, as indicated in the inset of Fig.\ \ref{fig5}.
This means that the scaling of the excitation size is weaker than
the size of the system. Hence, in the thermodynamic limit, 
(at least these) low-energy excitations will not cover a finite fraction
of the system. These results indicate that the energy landscape of NWP
is rather simple, i.e., dominate by one global minimum and close
local minima.

\section{Analytic approach: polymer in random media}
\label{sect:cavity}
The problem of negative-weight percolation has an interesting
correspondence with the problem of polymers in random media. We shall
show here that this correspondence can be used to study analytically
the percolation threshold on a random graph with a fixed connectivity.

Consider first a closed polymer of length $L$ described by a self-avoiding walk
$x=(x_1,\dots,x_L)$ on the graph $ G$, where $x_i\in V$ are $L$ distinct
vertices of
$G$ and each pair $(x_i,x_{i+1})$, as well as $(x_L,x_1)$, belongs to the set of edges $E$ of
$G$. The polymer is subject to a random potential $\omega$ on the edges of the
graph, so the energy of the polymer is $E_L(x)=\sum_{i=1}^L
\omega_{x_i,x_{i+1}}$ (we use the notation $x_{L+1}=x_1$). We suppose
that the values  of $\omega$ on the edges are iid quenched random
variables drawn from the distribution $P(\omega)$ given
by Eq.\ (\ref{eq:def:weights}).

We consider now a gas of such polymers, mutually avoiding, and we
denote by $M$ the total number of polymers present in the system. We
study this problem at equilibrium in a 
grand-canonical ensemble in which the inverse temperature is equal to
$\beta$ and the chemical potential is equal to
$\mu$. 
Note that in the remainder of this section, the inverse
temperature is denoted  by $\beta$. In order not to confuse it with
the order parameter exponent introduced previously 
in Eq.\ (\ref{eq:def:beta}), the latter one will be
referred to as \emph{critical exponent} $\beta$ in the remainder of this subsection. 
The partition function describing this system is 
\begin{eqnarray}
Z=\sum_{M=0}^\infty e^{\beta\mu M}\sum_{\{x\}}e^{-\beta E_M(x)}\ ,
\label{eq:zpol}
\end{eqnarray}
where $E_M=\sum_n E_{L_n}$ is the total energy of the various polymers
of lengths $L_1,L_2,\dots$.

A similar description of polymers in a solvent has been introduced
already in \cite{DeGennes75}. The study of homopolymers (without
disorder), shows that, in order to describe free polymers (in equilibrium with the solvent) the 
chemical potential has to be adjusted to a critical value $\mu_c$. 
This critical value corresponds to a phase transition 
between an infinitely diluted phase for $\mu<\mu_c$ and a dense 
phase with $\langle M \rangle /N>0$ (and a non-vanishing osmotic
pressure) for $\mu>\mu_c$. From Eq.\ (\ref{eq:zpol}), one finds that $\mu_c$
is equal to the canonical free energy density of a gas of polymers:
$\mu_c=-(1/(\beta M))\log(\sum_{\{x\}}e^{-\beta E_M(x)})$.
If the phase transition is continuous
the density on the coexistence line vanishes. Studies of interacting
heteropolymers have confirmed this statement \cite{MulMezMon,MulMezMon_long}. We
shall therefore develop a formalism allowing to compute the average
density of monomers on a point, $\psi$, as function of the inverse
temperature $\beta$, the chemical potential $\mu$ and the amount of
disorder $\rho$. For fixed value of $\beta$ and $\rho$, the critical
chemical potential $\mu$ is obtained as the largest value of $\mu$
where $\psi=0$. 

We conjecture that this problem of polymer in random medium is
equivalent to the NWP problem in the following sense. The set of polymers defines the loops
of NWP. Taking the zero temperature   ($\beta \to \infty$) limit, one
can study if the
ground-state energy of the set of polymers is negative (then the
corresponding polymer configuration is an admissible NWP), or not. The
phase transition of NWP then describes a transition between a
`percolating' phase
where the density $\psi$ of the ground state is positive, and a
non-percolating phase where it vanishes. Within our conjecture, we can
thus study the NWP phase transition as follows: we should find the
ground state of the polymer, at $\mu=\mu_c=0$: the condition
$\mu=\mu_c$ allows to study the polymer at equilibrium, the condition
$\mu_c=0$ finds the critical point of NWP where the ground state energy vanishes. 
We formulate this equivalence between NWP and polymers as a
conjecture for the following reason. In the polymer
approach we request that the energy of the full set of polymers be
negative, while NWP requests that each individual polymer have
negative energy. considering the thermodynamic limit where the length of the
polymers is large, and the fact that on a random regular graph almost
all loops are large since the typical size of closed circuits on such
a graph scale like $\log(N)$, it is reasonable to assume that the
distribution of energies of each polymer will follow a large deviation
principle, so that the request of negative energy for the full set
should be equivalent to the request of negative energy for each polymer.

We shall see below that the analytic solution of the
polymer problem with the cavity method is in good agreement with the
numerical results on the NWP. These results suggest that the
conjecture is correct, and that the NWP can be studied analytically
through this correspondence with  polymers.

\begin{figure}[t!]
\centerline{
\includegraphics[width=0.8\linewidth,angle=0]{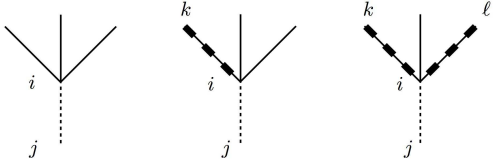}}
\caption{
 In a cavity graph where the edge $(ij)$ has been removed, there are
 three possible conformations of a polymer on site $i$. Left: no
 polymer. The probability of these configurations is $p^{(0)}_{i\to j}$. Center: The polymer arrives
 on $i$ from some other site $k$ (and will be forced to continue on the edge $(ij)$ when this
 edge will be put back in the graph). The probability of these
 configurations is $p^{(1)}_{i\to j}$. Right: The polymer goes through
 $i$, connecting two sites $k$ and $\ell$. The probability of these
 configurations is $p^{(2)}_{i\to j}$.
}
\label{fig:cav_pol}
\end{figure}  
The polymer problem can be studied using the cavity method as developed in
\cite{MulMezMon,MulMezMon_long}. Here, we shall use only the replica symmetric cavity
method as we have not found any evidence for sizeable replica symmetry
breaking effects. This is justified also by the results
for the low-energy excitations presented in section
\ref{sect:excitation}.
We consider a vertex $i$, we denote by $\partial i$
the set of vertices which are connected to $i$ by an edge, and we
study the cavity graph in which one edge $\{i,j\}$ (with $j\in\partial
i$) has
been removed. We consider the probabilities of the three possible configurations of
site $i$, as shown in Fig.\ \ref{fig:cav_pol}.
\begin{itemize}
\item The probability that there is no polymer on $i$ is denoted
  $p^{(0)}_{i\to j}$.
\item The probability that there is a polymer arriving on $i$ from one
  edge connecting $i$ to a vertex $k\in\partial i\setminus j$ (meaning
  that it will be forced to use the edge $\{i,j\}$). 
This probability is denoted
  $p^{(1)}_{i\to j}$.
\item The probability that the polymer goes through $i$ and uses two
  edges connecting $i$ to two  vertices $k,\ell$ in $\partial i \setminus j$. It is denoted
  $p^{(2)}_{i\to j}$.
\end{itemize}

The random regular graph is locally tree-like, which means that the
environment up to any finite distance from a generic point is a
regular tree. Using this property, which justifies the fact that
random 
variables on different branches of the cavity graph are independent
(within the replica symmetric hypothesis \cite{MezMon_book}),
one can show that these probabilities satisfy the following recursion
rules:
\begin{eqnarray}\nonumber
p^{(0)}_{i\to j} &=&C \prod_{m\in\partial i\setminus j}(p^{(0)}_{m\to
i}+p^{(2)}_{m\to i})\\ \nonumber
p^{(1)}_{i\to j} &=& C e^{\beta \mu}\sum_{k\in\partial i \setminus j}
p^{(1)}_{k\to i}e^{-\beta \omega_{ki}}\prod_{m\in\partial i \setminus j,k}(p^{(0)}_{m\to
i}+p^{(2)}_{m\to i})\\ \nonumber
p^{(2)}_{i\to j}&=&C e^{\beta \mu}\sum_{k<\ell\in\partial i\setminus j}
p^{(1)}_{k\to i}e^{-\beta \omega_{ki}} p^{(1)}_{\ell\to i} e^{-\beta
\omega_{\ell i}} \\
&&\prod_{m\in\partial i\setminus j,k,\ell}(p^{(0)}_{m\to
i}+p^{(2)}_{m\to i}) \ .
\end{eqnarray}

\begin{figure}[t!]
\centerline{\includegraphics[width=1.\linewidth]{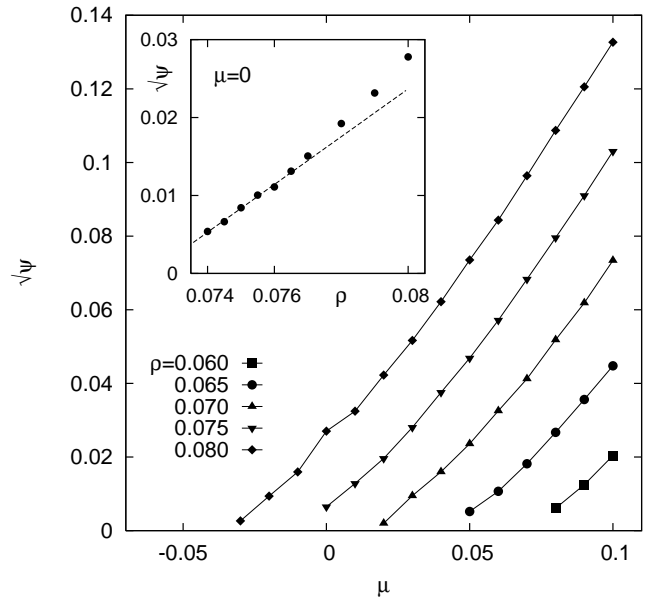}}
\caption{
 The average density of the polymer at $\beta=\infty$ has been computed 
by solving
Eq.\ (\ref{eq:qdeh}) for $Q(h)$  with the population dynamics method, with a
population size of $100000$ fields, and using the
resulting probability density in Eq.\ (\ref{eq:rhopop}). The plot shows
$\sqrt{\psi}$ as function of the chemical potential $\mu$. The various
curves correspond to different values of the strength of disorder
$\rho$: from right to left, $\rho= 0.06,0.065,\dots,0.080$.
The inset shows $\sqrt{\psi}$ versus $\rho$ for $\mu=0$.
}
\label{fig:density}

\end{figure}

The constant $C$ is a normalization constant, it is  fixed by the
  condition $p^{(0)}_{i\to j} +p^{(1)}_{i\to j} +p^{(2)}_{i\to j} =1 $.
The knowledge of the cavity probabilities $p^{(0)}_{i\to
  j},p^{(1)}_{i\to j},p^{(2)}_{i\to j}$ allows to obtain the various
physical quantities. For instance, the probability $\psi_i$ that site $i$ is
visited by a polymer is given by
\begin{eqnarray}
\psi_i=\frac{N_i}{D_i}\ ,
\end{eqnarray}
where
\begin{eqnarray}
N_i&=&e^{\beta \mu}\sum_{k<\ell\in\partial i}p^{(1)}_{k\to i}e^{-\beta \omega_{ki}} p^{(1)}_{\ell\to i} e^{-\beta
\omega_{\ell i}} \\
&&\prod_{m\in\partial i\setminus k,\ell}(p^{(0)}_{m\to
i}+p^{(2)}_{m\to i}) \ ,
\end{eqnarray}
and
\begin{eqnarray}
D_i= \prod_{m\in\partial i}(p^{(0)}_{m\to
i}+p^{(2)}_{m\to i}) +N_i\ .
\end{eqnarray}
It is possible to simplify the equations by a change of variables. Let
us introduce
\begin{eqnarray}
h_{ij} \equiv h_{i\to j}=  \frac{1}{\beta} \log \frac{p^{(1)}_{i\to j}
}{p^{(0)}_{i\to j} +p^{(2)}_{i\to j} } \ .
\end{eqnarray}
 One can write  the cavity
equations in terms only of the fields $h_{i\to j}$:
\begin{eqnarray}
h_{i\to j} =\frac{1}{\beta}\log\frac {
e^{\beta \mu}
\sum_{k\in\partial i \setminus j}
e^{\beta (h_{ki}-\omega_{ki})}
}
{
1+ e^{\beta \mu}
\sum_{k<\ell\in\partial i\setminus j}
e^{\beta (h_{ki}-\omega_{ki}+h_{\ell i}-\omega_{\ell i})} 
}\ .
\label{eq:cav_h}
\end{eqnarray}
These equations can be solved  by iteration, and their
solution gives access to various  properties of the polymer. The probability that a given site $i$ is occupied is given by
\begin{eqnarray}
\psi_i= \frac {
e^{\beta \mu}
\sum_{k<\ell\in\partial i}
e^{\beta (h_{ki}-\omega_{ki}+h_{\ell i}-\omega_{\ell i})} 
}
{
1+ e^{\beta \mu}
\sum_{k<\ell\in\partial i}
e^{\beta (h_{ki}-\omega_{ki}+h_{\ell i}-\omega_{\ell i})} 
}\ ,
\end{eqnarray}
and the probability that the edge $(ij)$ is occupied 
 is given by
\begin{eqnarray}
\psi_{(ij)}= \frac {
e^{\beta (h_{ij}+h_{ji}-\omega_{ij})} 
}
{
1+e^{\beta (h_{ij}+h_{ji}-\omega_{ij})} 
}\ .
\end{eqnarray}
The above equations allow for the study of one given instance of the
problem, i.e. one given realisation of the random graph and of the
potential. In the thermodynamic limit,  average properties like the average site
density $\psi=\frac{1}{N}\sum_i\psi_i$ are self-averaging (their
fluctuations from sample to sample go to zero in the large $N$
limit). These self-averaging properties can be studied using the
distribution of cavity fields $Q(h)$, which is the probability density that
the field $h_{i\to j}$ on an edge $(ij) \in\mathcal{E}$ chosen at
random uniformly in a randomly chosen sample is equal to $h$. The
cavity Eqs.\ (\ref{eq:cav_h}) imply that $Q(h)$ satisfies the
integral equation:
\begin{eqnarray}
\label{eq:qdeh}
Q(h)&=&\int\prod_{n=1}^K\left[Q(h_n)dh_n
  P(\omega_n)d\omega_n\right]\;\\
&\delta&\left(
h-\frac{1}{\beta}\log\frac {
e^{\beta \mu}
\sum_{n}
e^{\beta (h_{n}-\omega_{n})}
}
{
1+ e^{\beta \mu}
\sum_{m<n}
e^{\beta (h_{m}-\omega_{m}+h_{n}-\omega_{n})} 
}
\right)
\ , \nonumber
\end{eqnarray}
where $K=r-1$ is the number of neighbours of the root site in the
cavity graph, and the indices $m$ and $n$ run from $1$ to $K$. This integral
equation can be solved efficiently by the numerical approach of `population dynamics' described in
\cite{MezardParisi01}. The average density is then obtained as:
\begin{eqnarray}\nonumber
\psi= \int&&\prod_{n=1}^r\left[Q(h_n)dh_n
  P(\omega_n)d\omega_n\right]\;\\
&&\frac {
e^{\beta \mu}
\sum_{k<\ell}
e^{\beta (h_{k}-\omega_{k}+h_{\ell }-\omega_{\ell })} 
}
{
1+ e^{\beta \mu}
\sum_{k<\ell}
e^{\beta (h_{k}-\omega_{k}+h_{\ell }-\omega_{\ell })} 
}\ .
\label{eq:rhopop}
\end{eqnarray}
We have used this approach to study the polymer at zero temperature
(using the infinite $\beta$ limit of the Eqs.\
(\ref{eq:qdeh}),(\ref{eq:rhopop})) which is are easily written. 
Fig.\ \ref{fig:density} shows the average density computed by solving
Eq.\ (\ref{eq:qdeh}) for $Q(h)$  
with the population dynamics, and using the
resulting probability density in Eq.\ (\ref{eq:rhopop}). The data shows
that, for a given strength of disorder $\rho$, the average density $\psi$
vanishes at a critical value $\mu_c(\rho)$ of the chemical potential
$\mu$, with a quadratic behaviour:
\begin{eqnarray}
\psi\simeq A(\mu-\mu_c(\rho))^2\ .
\label{eq:x}
\end{eqnarray}
According to our above conjecture, the critical value $\rho_c$ of the
disorder strength is  obtained when the corresponding chemical
potential vanishes: $\mu_c(\rho_c)=0$. With this method one obtains the
prediction $\rho_c=0.072\pm 0.002$.
One also sees from Fig.\ \ref{fig:density},
by observing the shift of the curves when changeing the value
of $\rho$,
that $\mu_c(\rho)$ has a linear dependence on $\rho$ when $\rho\to
\rho_c$: 
\begin{eqnarray}
\mu_c(\rho)\simeq B (\rho-\rho_c)\ .
\label{eq:muc}
\end{eqnarray}
Combining Eqs.\ (\ref{eq:x}) and (\ref{eq:muc}) shows that the
density of the polymer at zero chemical potential grows quadratically
when $\rho>\rho_c$: $\psi(\mu=0)=A B^2 (\rho-\rho_c)^2$. 
This can also be seen directly by studying the $\mu=0$
case explicitly, see the inset of 
Fig.\ \ref{fig:density}.
This gives
according to Eq.\ (\ref{eq:def:beta}) the
value of the orderparameter exponent $\beta=2$.

These results show a good agreement between the analytical and numerical
approaches both for the value of $\rho_c$ and for the value of the critical exponent $\beta$.

\section{Conclusions  \label{sect:conclusions}}

We have presented an analytical and a numerical study of the NWP
problem in a mean-field case. 
The analytical study is based on a conjectured equivalence
with the problem of SAWs in random medium. The numerical study is based
on a mapping to a minimum-weight matching problem for which fast
algorithms exist. Both approaches yield results which are in
agreement, on the location of the phase transition, on the value of
critical exponents, and on the absence of any sizeable indications of
a glass phase: this is seen analytically in the fact that we do not
find any spin glass instability, and in the numerical study from the
sub-extensive scaling of the size of low-lying exitations. It is
interesting to note that the simulations of NWP using the
minimum-weight matching  turn out to be a very efficient numerical
approach to the difficult problem of polymers in random media.

To conclude, we summarize the numerical results found for the 
scaling exponents of NWP on regular random graphs with fixed 
connectivity $r\!=\!3$:
\begin{align*}
{\rm RRGs}:&&  \nu^*&=3.0(1) & \beta&=2.0(1)\\
&&  d_f &=2.1(1)   & \gamma&=-1.02(2) \,.
\end{align*}
These agree within error bars with the results found 
previously \cite{melchert2010a} for the NWP problem on 
$6d$ hypercubic lattice graphs:
\begin{align*}
6d:        &&  d\nu&=3.00(1) & \beta&=1.92(6)  \\
   && d_f&=2.00(1)  &  \gamma&=-0.99(3) \,. 
\end{align*}

These results further support an upper critical 
dimension $d_u\!=\!6$ for the NWP problem.

Via these and previous results, the static behavior of ordinary
NWP is now rather well understood. Since the NWP allows for a very
efficient numerical treatment, it would be quite rewarding to study
NWP on directed graphs or NWP on systems exhibiting correlated disorder.
Also it could be interesting to study the dynamic behavior using local
stochastic algorithms while comparing to the exact equilibrium behavior.

\begin{acknowledgments}
MM thanks the Humboldt foundation for its support, and the University
of Oldenburg for its hospitality. 
OM acknowledges financial support from the DFG (\emph{Deutsche Forschungsgemeinschaft})
under grant HA3169/3-1.
The simulations were performed at the GOLEM I cluster for scientific 
computing at the University of Oldenburg (Germany).
AKH thanks the LPTMS, Universit\'e Paris Sud (Orsay) for its
hospitality and financial support.
\end{acknowledgments}


\bibliography{literature_nwp_randGraph.bib}

\end{document}